\documentclass[aps,preprint]{revtex4}%
\usepackage{amsfonts}
\usepackage{amsmath}
\usepackage{amssymb}
\usepackage{graphicx}%
\begin{document}
\title[Hidden Momentum]{Interaction of a Point Charge and a Magnet: Comments on "Hidden Mechanical
Momentum Due to Hidden Nonelectromagnetic Forces"}
\author{Timothy H. Boyer}
\affiliation{Department of Physics, City College of the City University of New York, New
York, New York 10031}

\begin{abstract}
The interaction of a point charge and a magnetic moment (and by extension a
point charge and a solenoid) is explored within well-defined point-charge
magnetic-moment models where full calculations are possible. \ It is shown
explicitly how the "hidden mechanical momentum" is introduced by the "hidden"
external forces of constraint, requiring a prescribed response (through order
1/c$^{2})$ of the system to electromagnetic forces. \ These external forces
often go unmentioned in the textbook and research literature. \ The dependence
of "hidden mechanical momentum" upon detailed external (nonelectromagnetic)
forces may undermine the idea's usefulness in describing nature. \ Some
statements of dubious validity in the textbook literature are noted.

\end{abstract}
\maketitle

\section{Introduction}

Although the interaction of a point charge and magnet has been discussed
repeatedly for decades with some excellent analyses,\cite{SJ}\cite{CV}%
\cite{F}\cite{JCH} a convincing account of the interaction remains elusive,
particularly when there is relative motion of the objects. \ Furthermore, some
of the ideas based upon the assumption of stationary behavior have been
incorporated into the research and textbook literature with statements of
dubious validity. \ We explore this problem yet again. \ We consider two
familiar (point-charge) models for a magnetic moment which allow full
calculations for their interaction with a distant point charge. We also
consider stacking the magnetic moments to produce solenoids and so revisit the
interaction of a point charge and a constant-current solenoid. \ It is found
that the models require external forces of constraint which introduce linear
and angular momentum into the systems. \ Our discussion of simple, explicit
models appears to make a sacrifice in generality compared with the earlier
treatments. \ However, we believe that the advantage of the simple models is a
gain in clarity, particularly in understanding the changes in the magnetic
system and \ the role of the forces of constraint.

One version\cite{APV}\cite{V}\cite{H} for the analysis of the charge-magnet
interaction has been particularly influential in the recent literature. \ It
concludes that because of "hidden momentum" a new equation of motion is
required for (the center of energy of) a magnetic moment, and the associated
ideas have now entered the textbooks of classical electromagnetism\cite{G}%
\cite{J} in connection with the idea of "hidden mechanical momentum."
\ However, discussions of "hidden mechanical momentum" are often inaccurate,
superficial treatments of complex situations. \ "Hidden momentum" involves
linear momentum terms of order $1/c^{2},$ and discussions of "hidden momentum"
often suggest a form of behavior which is inaccurate even in nonrelativistic
physics, because the "hidden momentum" requires detailed (through order
$1/c^{2})$ forces of constraint which often go unmentioned in the textbook and
the research literature. \ In this article, we will not solve the complex
problem of the interaction of \ charges and magnets. \ Rather, within specific
(point-charge) models, we will explore the interaction in detail, noting the
conservation laws and the external forces associated with "hidden mechanical
momentum." \ 

\subsection{Outline of the Presentation}

We consider two models for a magnetic moment which are widely used in the
physics literature. \ In the first model, a magnetic moment is treated as a
charge moving with \textit{constant} \textit{speed} in a circular orbit plus a
second opposite charge at the center of the orbit. \ In the second model, a
magnetic moment is treated as a charge moving in a circular orbit of fixed
radius with an opposite charge at the center of the orbit, but now the speed
of the moving particle need not be constant. \ These two models are used to
explore the interaction of a point charge and a magnetic moment, and by
extension, the interaction of a point charge and a solenoid. \ 

First we treat the model where the magnetic moment charges move in circles
with constant speed. \ We start our discussion with a point charge outside a
constant-current solenoid, and then move to the discussion of a distant point
charge outside the orbit of a charged particle in uniform circular motion.
\ We find all the external forces on the systems and evaluate all the
conservation laws connected to these external forces. \ For these
constant-speed models, there is no "hidden momentum" in the systems. \ Next we
consider a magnetic moment where the speed of the charge is allowed to change
in response to electric forces tangential to its velocity. \ We note that (as
a theorem of mechanics) any point mass $m$ which is constrained to move in an
exact circle while subject to a constant perturbing force $\mathbf{F}$
acquires an average linear momentum $\left\langle \mathbf{p}_{m}%
^{mech}\right\rangle $ in order $1/c^{2}$ which is related to its angular
momentum $\mathbf{L}$ and the perturbing force $\mathbf{F}$ as $\left\langle
\mathbf{p}_{m}^{mech}\right\rangle =[1/(2mc^{2})]\mathbf{L\times F}$. \ This
is the "hidden mechanical momentum" which appears in the electromagnetism
textbooks (but not in the mechanics texts). \ We note that this "hidden
momentum" is related to the power flow delivered by the perturbing force, but
is introduced into the system by the impulse delivered by the external
centripetal forces\ of constraint. \ We also note that the constrained
response of the system which produces "hidden mechanical momentum" in a
charged-particle system will will necessarily create a speed-dependent
electric dipole moment for the system, which does not seem to be mentioned in
the literature. \ Finally we turn to a general discussion of the interaction
between charges and magnets. \ We note the historical context of the
discussion, comment on some statements of dubious validity in the textbook
literature, and place the problem in the context of previous and continuing
experimental work.

\section{Constant-Speed Model for Solenoids and Magnetic Moments}

\subsection{A Point Charge and a Constant-Current Solenoid}

When a point charge $q$\ is held at rest outside a very long solenoid, it is
claimed in some of the literature\cite{AR} that there are no forces exerted by
one system on the other. \ Now it is known from elementary electromagnetism
that a long, isolated, neutral solenoid has no electric or magnetic fields
outside its winding. It is suggested that therefore the solenoid exerts no
force on the external point charge $q$. \ However, the electric fields of the
point charge $q$ certainly exert forces on the charge carriers of the
solenoid. \ No one has ever done a convincing calculation of the response of
the solenoid charge carriers to the point charge fields, and therefore we do
not know the actual behavior of this system. \ However, we can discuss a
related model where the currents of the solenoid are held constant despite the
electric force due to the point charge fields experienced by the solenoid
charge carriers.\cite{B1973} \ This model requires that there are
nonelectromagnetic\ external forces on the charge carriers of the solenoid
which balance the electrical forces due to the point charge fields.

In order to make the situation as symmetrical as possible, the solenoid
currents are sometimes regarded as carried by opposite charges of equal
magnitude rotating in opposite directions with the cylindrical rings of
charged particles differing infinitesimally in radius.\cite{AR} \ The energy,
linear momentum, angular momentum, and energy-times-center-of-energy\cite{CE}
for our electromagnetic system of a point charge and constant-current solenoid
involve the sums of the mechanical and electromagnetic contributions. \ Since
this system is stationary in time, none of the conservation-related quantities
is changing in time. \ For simplicity of calculation here, we assume here that
the radius $r$ of the solenoid (with axis along the $z$-axis) is small
$r<<x_{q}$ compared to the distance $x_{q}$ from the solenoid axis to the
point charge $q$ (located along the $x$-axis at $\mathbf{x}_{q}=\widehat
{i}x_{q})$. \ Then the integral for the linear momentum $\mathbf{P}_{em}$ in
the electromagnetic field (associated with the overlap of the point charge
electric field $\mathbf{E}_{q}$ and the solenoid magnetic field $\mathbf{B}%
_{0})$ can be simplified by approximating the value of $\mathbf{E}%
_{q}\mathbf{(r)}$ as its value $\mathbf{E}_{q}\mathbf{(}0,0,z)$ on the axis of
the solenoid, giving%
\begin{align}
\mathbf{P}_{em}  &  =%
{\displaystyle\int}
d^{3}r\,\frac{1}{4\pi c}\mathbf{E}_{q}(\mathbf{r)}\times\mathbf{B}_{0}=%
{\displaystyle\int}
d^{3}r\,\frac{1}{4\pi c}\left(  \frac{q(\widehat{k}z-\widehat{i}x_{q}}%
{(z^{2}+x_{q}^{2})^{3/2}}\right)  \times\widehat{k}B_{0}\nonumber\\
&  =\widehat{j}\frac{qr^{2}B_{0}}{2cx_{q}}=\frac{q}{c}\mathbf{A(r}_{q})
\end{align}
Here $\mathbf{A(r}_{q})$ is the vector potential of the infinite solenoid in
the Coulomb gauge evaluated at the position of the external charge $q.$ \ Thus
the system of a point charge $q$ outside a constant-current solenoid contains
\textit{linear momentum} in the electromagnetic field. On the other hand,
there is no electromagnetic interaction \textit{energy} between the point
charge $q$ and the solenoid since the point charge has only an electric field
and the solenoid is electrically neutral. \ 

We notice that all the external forces $\mathbf{F}_{i}^{ext}$\ required to
keep the solenoid currents constant are applied to the charges $e_{i}$
carrying the currents of the solenoid. \ Since the positive and negative
charges (moving in opposite directions) differ only infinitesimally in
location, the nonelectromagnetic external force density and the mechanical
momentum density both vanish, and hence the sum of the external forces on the
system and the total mechanical momentum are both zero. \ Thus the
conservation law for the total system momentum $\mathbf{P=P}_{mech}%
+\mathbf{P}_{em}$ takes the form%

\begin{equation}%
{\displaystyle\sum\limits_{i}}
\mathbf{F}_{i}^{ext}=0=\frac{d\mathbf{P}}{dt}=\frac{d\mathbf{P}_{em}}{dt}%
\end{equation}
corresponding to constant linear momentum $\mathbf{P}=\mathbf{P}_{em}$\ for
the charge-solenoid system. \ When switching between the positive and negative
charges at a single spatial location, the velocities are reversed along with
the forces. \ Thus the power density due to the external forces does not
vanish. \ However, \ the power delivered by the external forces $\mathbf{F}%
_{i}^{ext}$ on the charges $e_{i}$ reverses sign under reflection in the
$xz$-plane, so that the total power delivered by the external forces is zero%
\begin{equation}%
{\displaystyle\sum\limits_{i}}
\mathbf{F}_{i}^{ext}\cdot\mathbf{v}_{i}=0=\frac{dU}{dt}%
\end{equation}
corresponding to constant energy $U$ for the system. \ On the other hand, the
$y$-component of displacement also changes sign under reflection in the
$xz$-plane so that the power-weighted displacement of the system is not zero.
\ Indeed, we must have\cite{B5}%
\begin{equation}%
{\displaystyle\sum\limits_{i}}
(\mathbf{F}_{i}^{ext}\cdot\mathbf{v}_{i})\mathbf{r}_{i}=\frac
{d(U\overrightarrow{X})}{dt}-c^{2}\mathbf{P}%
\end{equation}
which becomes $%
{\displaystyle\sum}
(\mathbf{F}_{i}^{ext}\cdot\mathbf{v}_{i})\mathbf{r}_{i}=-c^{2}\mathbf{P}_{em}$
since neither the energy $U$ nor the center of energy $\overrightarrow{X}$ is
changing with time so that ($d/dt)(U_{em}\overrightarrow{X})=0,$ and
$\mathbf{P}_{mech}=0$ so that $-c^{2}\mathbf{P}=-c^{2}\mathbf{P}_{em}.$
\ Later we will calculated explicitly the left-hand side of Eq. (4).
\ However, here we merely emphasize that for this constant-current case, the
electromagnetic field momentum must be associated with the power densities of
nonrelativistic external forces of constraint. The electromagnetic field
momentum is directly related to the introduction of power in the
electromagnetic system at one spatial location and its removal at another
location. \ 

\subsection{Magnetic Moment Modeled as a Charge with Constant Angular
Velocity}

In order to simplify the analysis for the interaction of a point charge and a
magnet, we consider a solenoid as a stack of magnetic dipoles, and treat the
interaction of a point charge and a magnetic dipole. \ Our model of a magnetic
dipole is simplified further by considering only one moving charge and one
stationary charge of opposite sign. \ Specifically, our system consists of a
distant point charge $q$ located on the $x$-axis $\mathbf{r}_{q}=\widehat
{i}x_{q}$, a negative charge $-e$ along the $z$-axis at $z_{e},$ and a
particle of positive charge $e$ and mass $m$\ in uniform circular motion in
the plane $z=z_{e}$ with speed $v_{e}=\omega_{0}r$, the orbit being of radius
$r$ and centered on the $z$-axis. \ The combination of charges $-e$ and $+e$
is electrically neutral and has a magnetic moment given by\cite{J2}%
\begin{equation}
\overrightarrow{\mu}=\widehat{k}\frac{erv_{e}}{2c}%
\end{equation}
In this case the charge $e$ is moving with constant angular velocity as%
\begin{equation}
\mathbf{r}_{e}(t)=r[\widehat{i}\cos(\omega_{0}t)+\widehat{j}\sin(\omega
_{0}t)]+\widehat{k}z_{e}%
\end{equation}%
\begin{equation}
\mathbf{v}_{e}(t)=r\omega_{0}[-\widehat{i}\sin(\omega_{0}t)+\widehat{j}%
\cos(\omega_{0}t)]
\end{equation}%
\begin{equation}
\mathbf{a}_{e}=-r\omega_{0}^{2}[\widehat{i}\cos(\omega_{0}t)+\widehat{j}%
\sin(\omega_{0}t)]
\end{equation}
Now we will calculate the momentum, energy, and angular momentum for this
system consisting of point masses and electromagnetic fields. \ We will then
average over the periodic motion to obtain average values for all the
quantities. \ Once again, the constraints imposed upon the system must be
provided by nonelectromagnetic external forces $\mathbf{F}_{e}^{ext},$
$\mathbf{F}_{-e,}^{ext}$ and $\mathbf{F}_{q}^{ext}$ acting on respectively the
charge $e$, the charge $-e$, and the charge $q$. \ 

\subsection{Conservation-Related Quantities for the Magnetic Moment}

The values (mechanical plus electromagnetic) for the system energy, linear
momentum, angular momentum, and energy times center of energy can all be
obtained through order $1/c^{2}$ by integrating over the field densities.
\ This leads\cite{PA} to the values which are obtained from the Darwin
Lagrangian.\cite{CV} \ We will time-average over these values, retaining only
through first-order terms in $r/x_{q}.$ \ Then the displacement from the
charge $q$ to the moving charge $e$ is given by $\mathbf{r}_{e}-\mathbf{r}%
_{q}=\widehat{i}[r\cos(\omega_{0}t)-x_{q}]+\widehat{j}r\sin\left(  \omega
_{0}t\right)  +\widehat{k}z_{e}$ with the distance approximated as $\left\vert
\mathbf{r}_{e}-\mathbf{r}_{q}\right\vert ^{n}=\{x_{q}^{2}-2rx_{q}\cos
(\omega_{0}t)+r^{2}+z_{e}^{2}\}^{n/2}=(x_{q}^{2}+z_{e}^{2})^{n/2}%
\{1-[nrx_{q}\cos(\omega_{0}t)]/(x_{q}^{2}+z_{e}^{2})\}.$ \ The time-averages
(indicated by brackets $\left\langle {}\right\rangle )$ give $\left\langle
\sin^{2}(\omega_{0}t)\right\rangle =\left\langle \cos^{2}(\omega
_{0}t)\right\rangle =1/2$, and $\left\langle \sin(\omega_{0}t)\cos(\omega
_{0}t)\right\rangle =$ $0.$

The total energy of the system includes mechanical plus electromagnetic energy
and gives an average value%
\begin{align}
\left\langle U\right\rangle  &  =\left\langle m\gamma_{e}c^{2}-\frac{e^{2}}%
{r}+\frac{eq}{[x_{q}^{2}-2r\cos(\omega_{0}t)+r^{2}+z_{e}^{2}]^{1/2}}-\frac
{eq}{[x_{q}^{2}+z_{e}^{2}]^{1/2}}+M_{-e}c^{2}+M_{q}c^{2}\right\rangle
\nonumber\\
&  =m\gamma_{e}c^{2}-\frac{e^{2}}{r}+M_{-e}c^{2}+M_{q}c^{2}%
\end{align}
where we have written $\gamma_{e}=(1-v_{e}^{2}/c^{2})^{-1/2}$ with
$v_{e}=\omega_{0}r.$ Here we are ignoring electric quadrupole terms of order
$r^{2}/x_{q}^{2}$. \ We notice that there is no time-average interaction
energy between the stationary charge $q$ and the magnetic dipole.

Using the expression for the velocity of charge $e$ given in Eq. (7), the
linear momentum is\cite{PA}\cite{CV}%
\begin{align}
\left\langle \mathbf{P}_{\mu}\right\rangle  &  =\left\langle m\gamma
_{e}\mathbf{v}_{e}+\frac{qe}{2c^{2}}\left(  \frac{\mathbf{v}_{e}}{r_{qe}%
}+\frac{(\mathbf{v}_{e}\cdot\mathbf{r}_{qe})\mathbf{r}_{qe}}{r_{qe}^{3}%
}\right)  \right\rangle \nonumber\\
&  \mathbf{=}\frac{_{av}\mathbf{/}}{\backslash}m\gamma_{e}\mathbf{v}%
_{e}\mathbf{+}\frac{qe}{2c^{2}}\{\frac{\mathbf{v}_{e}}{[x_{q}^{2}-2rx_{q}%
\cos(\omega_{0}t)+r^{2}+z_{e}^{2}]^{1/2}}\nonumber\\
&  +\frac{\mathbf{v}_{e}\mathbf{\cdot\{}\widehat{i}[r\cos(\omega_{0}%
t)-x_{q}]+\widehat{j}r\sin(\omega_{0}t)+\widehat{k}z_{e}\}}{[x_{q}^{2}%
-2rx_{q}\cos(\omega_{0}t)+r^{2}+z_{e}^{2}]^{3/2}}\{\widehat{i}[r\cos
(\omega_{0}t)-x_{q}]+\widehat{jr}\sin(\omega_{0}t)+\widehat{k}z_{e}%
\}\}\frac{\backslash_{av}}{/}\nonumber\\
&  =\frac{_{av}/}{\backslash}\frac{qe}{2c^{2}}\{\frac{v_{e}[-\widehat{i}%
\sin(\omega_{0}t)+\widehat{j}\cos(\omega_{0}t)]}{[x_{q}^{2}+z_{e}^{2}]^{1/2}%
}\left(  1+\frac{rx_{q}\cos(\omega_{0}t)}{x_{q}^{2}+z_{e}^{2}}\right)
\nonumber\\
&  +\frac{-v_{e}\sin(\omega_{0}t)[r\cos(\omega_{0}t)-x_{q}]+v_{e}\cos
(\omega_{0}t)r\sin(\omega_{0}t)}{[x_{q}^{2}+z_{e}^{2}]^{3/2}}\left(
1+\frac{3rx_{q}\cos(\omega_{0}t)}{x_{q}^{2}+z_{e}^{2}}\right) \nonumber\\
&  \times\{\widehat{i}[r\cos(\omega_{0}t)-x_{q}]+\widehat{jr}\sin(\omega
_{0}t)+\widehat{k}z_{e}\}\}\frac{\backslash_{av}}{/}\nonumber\\
&  =\widehat{j}\frac{qerv_{e}x_{q}}{2c^{2}[x_{q}^{2}+z_{e}^{2}]^{3/2}%
}=\widehat{j}\frac{q\mu x_{q}}{c[x_{q}^{2}+z_{e}^{2}]^{3/2}}=\frac{1}%
{c}\mathbf{E}_{q}\mathbf{(}0,0,z_{e})\times\overrightarrow{\mu}%
\end{align}
where $\overrightarrow{\mu}=\widehat{k}erv_{e}/(2c)$ is the magnetic moment
produced by the moving charge $e,$ and $\mathbf{E}_{q}\mathbf{(}0,0,z_{e})$ is
the electric field of the charge $q$ evaluated at the position of the magnetic
moment. \ Since the charge $e$ moves with uniform circular motion and so has
no average linear momentum, the average system momentum is contained entirely
in the electromagnetic field. \ 

The angular momentum about the origin can also be evaluated using the
calculation above since we recognize the same average as was needed in Eq.
(10). \ The system angular momentum contains both the mechanical angular
momentum of the orbiting charge $e$ and the electromagnetic field angular
momentum\cite{PA}%
\begin{align}
\left\langle \mathbf{L}\right\rangle  &  =\left\langle \mathbf{r}_{e}\times
m\gamma\mathbf{v}_{e}\mathbf{+}\frac{qe}{2c^{2}}\widehat{i}x_{q}\times\left(
\frac{\mathbf{v}_{e}}{r_{qe}}+\frac{(\mathbf{v}_{e}\cdot\mathbf{r}%
_{qe})\mathbf{r}_{qe}}{r_{qe}^{3}}\right)  \right\rangle \nonumber\\
&  =\widehat{k}m\gamma rv+\widehat{i}x_{q}\times\widehat{j}\frac{qerv_{e}%
x_{q}}{2c^{2}[x_{q}^{2}+z_{e}^{2}]^{3/2}}\nonumber\\
&  =\widehat{k}m\gamma_{e}rv_{e}+\widehat{k}\frac{qerv_{e}x_{q}^{2}}%
{2c^{2}[x_{q}^{2}+z_{e}^{2}]^{3/2}}=\widehat{k}m\gamma rv_{e}+\widehat{k}%
\frac{\mu qx_{q}^{2}}{c[x_{q}^{2}+z_{e}^{2}]^{3/2}}\nonumber\\
&  =\mathbf{L}_{mech}+\mathbf{r}_{q}\times\left(  \frac{1}{c}\mathbf{E}%
_{q}\mathbf{(}0,0,z_{e})\times\overrightarrow{\mu}\right)
\end{align}

\subsection{Conservation-Related Quantities for a Solenoid}

From the results for the conservation-related quantities for a point charge
and a constant-speed magnetic moment, we can use integration to obtain the
corresponding quantity for the system of a point charge and a constant-current
solenoid. \ In the present model calculation, only one set of charges carries
the solenoid currents, in contrast to the discussion above of a solenoid with
moving positive and negative charge carriers; however, the electromagnetic
conservation-related quantities depend only upon the fields and are unchanged.
\ The average energy in Eq. (9) shows no average interaction energy between a
point charge and a magnetic moment, and we expect none between a point charge
and a constant-current solenoid. \ The linear momentum associated with the
interaction of a point charge and a solenoid can be obtained by integrating
Eq. (10) over a stack of magnetic moments using the replacement\ $erv_{e}%
/2c=\mu\rightarrow dz\,[\pi r^{2}B_{0}/(4\pi)]$ since $[\pi r^{2}B_{0}%
/(4\pi)]$ is the dipole moment per unit length for a solenoid. \ Thus the
momentum in the point charge-solenoid system is%
\begin{align}
\mathbf{P}_{\mu\rightarrow sol}  &  =\widehat{j}%
{\displaystyle\int\limits_{-\infty}^{\infty}}
dz\,\frac{qx_{q}}{c[x_{q}^{2}+z_{e}^{2}]^{3/2}}\left(  \frac{\pi r^{2}B_{0}%
}{4\pi}\right) \nonumber\\
&  =\widehat{j}\frac{qr^{2}B_{0}}{2cx_{q}}=\frac{q}{c}\mathbf{A}%
(\mathbf{r}_{q})
\end{align}
which is identical with the result obtained earlier in Eq. (1).

The angular momentum of the interacting point charge and constant-current
solenoid of finite length $L$ can also be obtained by integrating over Eq.
(11) for the angular momentum of the interacting point charge and magnetic
moment. \ The mechanical angular momentum $\widehat{k}m\gamma_{e}rv_{e}$
associated with the particle in uniform circular motion does not reflect the
charge-solenoid interaction and will diverge for an infinite solenoid. \ The
contribution from the angular momentum in the electromagnetic field for a
solenoid of length $L$ is given by the integral%
\begin{align}
\mathbf{L}_{\mu\rightarrow sol}^{(L)}  &  =\widehat{k}%
{\displaystyle\int\limits_{-L/2}^{L/2}}
dz\,\frac{qx_{q}^{2}}{c[x_{q}^{2}+z^{2}]^{3/2}}\left(  \frac{\pi r^{2}B_{0}%
}{4\pi}\right) \nonumber\\
&  =\widehat{k}\frac{qr^{2}B_{0}}{2c}\frac{L/2}{[x_{q}^{2}+(L/2)^{2}]^{1/2}}%
\end{align}
In the limit of an infinite solenoid $L\rightarrow\infty$, the field angular
momentum $\mathbf{L}_{\mu\rightarrow sol}$\ in Eq. (13) becomes
\begin{equation}
\mathbf{L}_{\mu\rightarrow sol}=\overrightarrow{k}qr^{2}B_{0}/(2c)=\mathbf{r}%
_{q}\times\mathbf{A}(\mathbf{r}_{q})
\end{equation}
and is independent of the separation $x_{q}$ of the point charge and the
solenoid\cite{AngMom}. \ Any stack of magnetic moments of \textit{finite}
length, as in Eq. (13), will have an electromagnetic angular momentum which
depends upon the distance $x_{q}$ from the solenoid to the charge $q,$ and
which vanishes when the point charge $q$ is removed infinitely far from the
stack of magnetic moments.

\subsection{External Forces of Constraint for the Constant-Speed Magnetic
Moment}

The quantities obtained in Eqs. (9)-(13) are those traditionally obtained in
the textbooks and the research literature. \ However, the literature sometimes
makes no reference to the role of the nonelectromagnetic external forces
acting on the charged particles of our electromagnetic system. \ Nevertheless,
the external forces of constraint are absolutely crucial in producing these
conservation-related quantities. \ Accordingly, we now wish to investigate the
forces $\mathbf{F}_{e}^{ext},$ $\mathbf{F}_{-e,}^{ext}$ and $\mathbf{F}%
_{q}^{ext}$.

The equation of motion for the charge $e$ in the magnetic moment of Eq. (5) is
given by $(d/dt)(m\gamma_{e}\mathbf{v}_{e})=\mathbf{F}_{e}^{ext}%
+e\mathbf{E}_{-e}(\mathbf{r}_{e})+e\mathbf{E}_{q}(\mathbf{r}_{e})$ which
becomes for uniform circular motion
\begin{equation}
-[\widehat{i}\cos(\omega_{0}t)+\widehat{j}\sin(\omega_{0}t)]m\gamma_{e}%
\omega_{0}^{2}r=\mathbf{F}_{e}^{ext}-[\widehat{i}\cos(\omega_{0}t)+\widehat
{j}\sin(\omega_{0}t)]\frac{e^{2}}{r^{2}}+\frac{eq(-\widehat{i}x_{q}%
+\overrightarrow{k}z_{e})}{(z_{e}^{2}+x_{q}^{2})^{3/2}}%
\end{equation}
where we have written $\gamma_{e}=(1-v_{e}^{2}/c^{2})^{-1/2}$ with
$v_{e}=\omega_{0}r,$ and have approximated the electric field due to $q$ as
its value at the center of the orbit since $r<<x_{q}.$ \ We can solve for
$\mathbf{F}_{e}^{ext}$ in Eq. (15) and then average over a period so as to
obtain $\left\langle \mathbf{F}_{e}^{ext}\right\rangle ,$ $\left\langle
\mathbf{F}_{e}^{ext}\cdot\mathbf{v}_{e}\right\rangle ,$ $\left\langle
\mathbf{r}_{e}\times\mathbf{F}_{e}^{ext}\right\rangle $ and $\left\langle
(\mathbf{F}_{e}^{ext}\cdot\mathbf{v}_{e})\mathbf{r}_{e}\right\rangle .$ We
find \
\begin{equation}
\left\langle \mathbf{F}_{e}^{ext}\right\rangle =\frac{eq(\widehat{i}%
x_{q}-\widehat{k}z_{e})}{(z_{e}^{2}+x_{q}^{2})^{3/2}}%
\end{equation}%
\begin{equation}
\left\langle \mathbf{F}_{e}\cdot\mathbf{v}_{e}\right\rangle =0
\end{equation}%
\begin{equation}
\left\langle \mathbf{r}_{e}\times\mathbf{F}_{e}^{ext}\right\rangle =0
\end{equation}
and%
\begin{align}
\left\langle (\mathbf{F}_{e}^{ext}\cdot\mathbf{v}_{e})\mathbf{r}%
_{e}\right\rangle  &  =_{av}/\left[  \left(  \widehat{i}\frac{eqx_{q}}%
{(z_{e}^{2}+x_{q}^{2})^{3/2}}\right)  \cdot\lbrack-\widehat{i}\sin(\omega
_{0}t)+\widehat{j}\cos(\omega_{0}t)]v\right] \nonumber\\
&  \times\{r[\widehat{i}\cos(\omega_{0}t)+\widehat{j}\sin(\omega
_{0}t)]+\widehat{k}z_{e}\}\backslash_{av}\nonumber\\
&  =-\widehat{j}\frac{eqx_{q}rv_{e}}{2(z_{e}^{2}+x_{q}^{2})^{3/2}}%
\end{align}
The average force on the central negative charge is simply that due to the
charge $q$%
\begin{equation}
\left\langle \mathbf{F}_{-e}^{ext}\right\rangle =\frac{-eq(\widehat{i}%
x_{q}-\widehat{k}z_{e})}{(z^{2}+x_{q}^{2})^{3/2}}%
\end{equation}

With these results, we can evaluate the various conservation laws. \ The net
external force vanishes for the magnetic moment system of $\ $charges $e$ and
$-e,$ $\left\langle \mathbf{F}_{e}^{ext}\right\rangle +\left\langle
\mathbf{F}_{-e}^{ext}\right\rangle =0.$ \ The average force on the charge $q$
vanishes $\left\langle \mathbf{F}_{q}^{ext}\right\rangle =0,$ when we neglect
the average quadrupole moment of the system of charges $e$ and $-e.$ \ Thus
the average of the sum of the external forces vanishes $\left\langle
\mathbf{F}_{e}^{ext}\right\rangle +\left\langle \mathbf{F}_{-e}^{ext}%
\right\rangle +\left\langle \mathbf{F}_{q}^{ext}\right\rangle =0,$ and the
average system momentum given in Eq, (10) is a constant in time. \ The average
power delivered by the external forces also vanishes $\left\langle
\mathbf{F}_{e}^{ext}\cdot\mathbf{v}_{e}\right\rangle +\left\langle
\mathbf{F}_{-e}^{ext}\cdot\mathbf{v}_{-e}\right\rangle +\left\langle
\mathbf{F}_{q}^{ext}\cdot\mathbf{v}_{q}\right\rangle =\left\langle
\mathbf{F}_{e}^{ext}\cdot\mathbf{v}_{e}\right\rangle +\left\langle
\mathbf{F}_{-e}^{ext}\cdot0\right\rangle +\left\langle \mathbf{F}_{q}%
^{ext}\cdot0\right\rangle =0,$ and the system energy is constant in time.
\ The average total torque about the origin supplied by the sum of the forces
vanishes $\left\langle \mathbf{r}_{e}\times\mathbf{F}_{e}^{ext}\right\rangle
+\left\langle \mathbf{r}_{-e}\times\mathbf{F}_{-e}^{ext}\right\rangle
+\left\langle \mathbf{r}_{q}\times\mathbf{F}_{q}^{ext}\right\rangle
=\left\langle \mathbf{r}_{e}\times\mathbf{F}_{e}^{ext}\right\rangle
+\left\langle 0\times\mathbf{F}_{-e}^{ext}\right\rangle +\left\langle
\mathbf{r}_{q}\times0\right\rangle =0,$ and the average total angular momentum
is constant in time. \ However, from Eq. (19), the power-weighted displacement
$\left\langle (\mathbf{F}_{e}^{ext}\cdot\mathbf{v}_{e})\mathbf{r}%
_{e}\right\rangle +\left\langle (\mathbf{F}_{-e}^{ext}\cdot\mathbf{v}%
_{-e})\mathbf{r}_{-e}\right\rangle +\left\langle (\mathbf{F}_{q}^{ext}%
\cdot\mathbf{v}_{q})\mathbf{r}_{q}\right\rangle =\left\langle (\mathbf{F}%
_{e}^{ext}\cdot\mathbf{v}_{e})\mathbf{r}_{e}\right\rangle +\left\langle
(\mathbf{F}_{-e}^{ext}\cdot0)0\right\rangle +\left\langle (\mathbf{F}%
_{q}^{ext}\cdot0)\mathbf{r}_{q}\right\rangle =-\widehat{j}eqx_{q}%
rv_{e}/[2(z_{e}^{2}+x_{q}^{2})^{3/2}]$ is not zero and is indeed equal to
$c^{2}$ times the negative of the system linear momentum in Eq. (10), as
required by the time average of the relation in Eq. (4).

\subsection{External Forces of Constraint for a Constant-Current Solenoid}

From the expressions obtained in Eqs. (16)-(20), we can obtain the
corresponding expressions for a point charge $q$ interacting with a
constant-current solenoid. \ Thus we can evaluate $%
{\displaystyle\sum}
(\mathbf{F}_{i}^{ext}\cdot\mathbf{v}_{i})\mathbf{r}_{i}$ by integrating the
time average in Eq. (19) over the length of the solenoid. \ The magnetic
moment is replaced by $dz$ times the magnetic moment per unit length
$\mu=erv/(2c)\rightarrow dz\,\pi r^{2}B_{0}/(4\pi)$%
\begin{align}%
{\displaystyle\sum_{i}}
(\mathbf{F}_{i}^{ext}\cdot\mathbf{v}_{i})\mathbf{r}_{i}  &  =-\widehat{j}%
{\displaystyle\int\limits_{-\infty}^{\infty}}
dz\,\frac{c\pi r^{2}B_{0}}{4\pi}\frac{qx_{q}}{(z^{2}+x_{q}^{2})^{3/2}%
}\nonumber\\
&  =-\widehat{j}\frac{cr^{2}B_{0}qx_{q}}{4}\left(  \frac{z}{x_{q}^{2}%
(z^{2}+x_{q}^{2})^{1/2}}\right)  _{-\infty}^{\infty}\nonumber\\
&  =-\widehat{j}\frac{cr^{2}B_{0}q}{2x_{q}}%
\end{align}
From Eqs. (1) and (21), we find that indeed equation (4) holds true for the
constant-current solenoid. \ It seems striking that work done by
nonrelativistic external forces in Eq. (21) is related directly to the
relativistic expression (1) for the linear momentum in the associated
electromagnetic fields.

\subsection{Quasistatic Changes for the Point Charge Near the Magnetic Moment}

The connection between the nonelectromagnetic external forces of constraint
and the electromagnetic field momentum and angular momentum can be made more
emphatic by bringing the charge $q$ quasistatically along the $x$-axis from
spatial infinity up to its final position $\mathbf{r}_{q}=\widehat{i}x_{q}.$
\ During the quasistatic change, the external nonelectromagnetic forces must
provide the rates of change for the system energy, linear momentum, and
angular momentum. \ The magnetic forces associated with the moving charges $e$
and $q$ provide nonvanishing impulses when averaged over the uniform circular
motion of the charge $e$ of the magnetic moment. \ The average external forces
of constraint can be evaluated alternatively as the time average over the
external forces required to maintain the uniform circular motion of the charge
$e$ and the quasistatic motion of the charge $q$ in the presence of the
magnetic forces, or, more simply, as the external forces needed to balance the
magnetic forces on the magnetic moment $\overrightarrow{\mu}$ and on the
charge $q.$ \ We will present the simpler calculations here. \ 

When the charge $q$ is moved quasistatically, it has a velocity $\mathbf{v}%
_{q}=\widehat{i}v_{q}=\widehat{i}(dx_{q}/dt)$ which may be taken as small as
desired. \ The magnetic moment $\overrightarrow{\mu}=\widehat{k}\mu$ creates a
magnetic field $\mathbf{B}_{\mu}$ which places a magnetic force $\mathbf{F}%
_{q}^{mag}$ on the charge $q.$ \ This force on $q$ must be balanced by a
nonelectromagnetic external force $\Delta\mathbf{F}_{q}^{ext-mag}%
=-\mathbf{F}_{q}^{mag}$ giving%
\begin{align}
-\Delta\mathbf{F}_{q}^{ext-mag}  &  =\mathbf{F}_{q}^{mag}=q(\mathbf{v}%
_{q}/c)\times\mathbf{B}_{\mu}(\mathbf{r}_{q})\nonumber\\
&  =\frac{q}{c}\frac{dx_{q}}{dt}\widehat{i}\times\left(  \frac{3[\mu
\widehat{k}\cdot(\widehat{i}x_{q}-\widehat{k}z_{e})](\widehat{i}x_{q}%
-\widehat{k}z_{e})}{(x_{q}^{2}+z_{e}^{2})^{5/2}}-\frac{\mu\widehat{k}}%
{(x_{q}^{2}+z_{e}^{2})^{3/2}}\right) \nonumber\\
&  =-\widehat{j}\frac{3q\mu z_{e}^{2}}{c(x_{q}^{2}+z_{e}^{2})^{5/2}}%
\frac{dx_{q}}{dt}+\widehat{j}\frac{q\mu}{c(x_{q}^{2}+z_{e}^{2})^{3/2}}%
\frac{dx_{q}}{dt}%
\end{align}
\ The impulse $\mathbf{I}_{q}^{ext}$ delivered by the external force when the
charge $q$ is moved from spatial infinity to the position $\mathbf{r}%
_{q}=\widehat{i}x_{q}$ follows from Eq. (22) as%
\begin{align}
\mathbf{I}_{q}^{ext}  &  =%
{\displaystyle\int}
dt\,\Delta\mathbf{F}_{q}^{ext-mag}=-\widehat{j}\frac{q\mu}{c}%
{\displaystyle\int}
dt\,\frac{dx_{q}}{dt}\left(  -\frac{3z_{e}^{2}}{(x_{q}^{2}+z_{e}^{2})^{5/2}%
}+\frac{1}{(x_{q}^{2}+z_{e}^{2})^{3/2}}\right) \nonumber\\
&  =-\widehat{j}\frac{q\mu}{c}%
{\displaystyle\int\limits_{\infty}^{x_{q}}}
dx\left(  -\frac{3z_{e}^{2}}{(x^{2}+z_{e}^{2})^{5/2}}+\frac{1}{(x^{2}%
+z_{e}^{2})^{3/2}}\right) \nonumber\\
&  =\widehat{j}\frac{q\mu}{c}\left[  \frac{x_{q}}{(x_{q}^{2}+z_{e}^{2})^{3/2}%
}+\frac{1}{z_{e}^{2}}\left(  \frac{x_{q}}{(x_{q}^{2}+z_{e}^{2})^{1/2}%
}-1\right)  \right]
\end{align}
Furthermore, when the charge $q$ is moved quasistatically, it generates a
magnetic field $\mathbf{B}_{q}$ which acts on the magnetic moment
$\overrightarrow{\mu}$ with a force $\mathbf{F}_{\mu}^{mag}$. \ This magnetic
force on $\mu$ (located at $\mathbf{r}_{\mu}=\widehat{k}z_{e})$ must be
balance by a nonelectromagnetic external force $\Delta\mathbf{F}_{\mu
}^{ext-mag}=-\mathbf{F}_{\mu}^{mag},$%
\begin{align}
-\Delta\mathbf{F}_{\mu}^{ext-mag}  &  =\mathbf{F}_{\mu}^{mag}=\nabla
\lbrack\overrightarrow{\mu}\cdot\mathbf{B}_{q}(\mathbf{r)]}_{\mathbf{r}_{\mu}%
}\nonumber\\
&  =\nabla\left[  \mu\widehat{k}\cdot\left(  q\frac{\mathbf{v}_{q}}{c}%
\times\frac{\widehat{i}(x-x_{q})+\widehat{j}y+\widehat{k}z}{[(x-x_{q}%
)^{2}+y^{2}+z^{2}]^{3/2}}\right)  \right]  _{\mathbf{r}_{\mu}}\nonumber\\
&  =\nabla\left[  \frac{q\mu y}{c[(x-x_{q})^{2}+y^{2}+z^{2}]^{3/2}}%
\frac{dx_{q}}{dt}\right]  _{\mathbf{r}_{\mu}}=\widehat{j}\frac{q\mu}%
{c(z_{e}^{2}+x_{q}^{2})^{3/2}}\frac{dx_{q}}{dt}%
\end{align}
\ The impulse $\mathbf{I}_{\mu}^{ext}$ delivered by the external force when
the charge $q$ is moved from spatial infinity to the position $\mathbf{r}%
_{q}=\widehat{i}x_{q}$ follows from Eq. (24) as%
\begin{align}
\mathbf{I}_{\mu}^{ext}  &  =%
{\displaystyle\int}
dt\,\Delta\mathbf{F}_{q\mu}^{ext-mag}=-\widehat{j}\frac{q\mu}{c}%
{\displaystyle\int}
dt\,\frac{dx_{q}}{dt}\frac{1}{(z_{e}^{2}+x_{q}^{2})^{3/2}}\nonumber\\
&  =-\widehat{j}\frac{q\mu}{c}%
{\displaystyle\int\limits_{\infty}^{x_{q}}}
dx\,\frac{1}{(z_{e}^{2}+x^{2})^{3/2}}=-\widehat{j}\frac{1}{z_{e}^{2}}\left(
\frac{x_{q}}{(x_{q}^{2}+z_{e}^{2})^{1/2}}-1\right)
\end{align}
If we add together the results of Eqs. (23) and (25) to obtain the total
impulse delivered to the electromagnetic system of point charge $q$ and
magnetic moment $\overrightarrow{\mu},$ we find%
\begin{equation}
\mathbf{I}_{total}^{ext}=\mathbf{I}_{q}^{ext}+\mathbf{I}_{\mu}^{ext}%
=\widehat{j}\frac{q\mu}{c}\frac{x_{q}}{(x_{q}^{2}+z_{e}^{2})^{3/2}}=\frac
{1}{c}\mathbf{E}_{q}\mathbf{(}0,0,z_{e})\times\overrightarrow{\mu}%
\end{equation}
which is exactly the average electromagnetic field momentum calculated in Eq.
(10). \ Thus for quasistatic charge motions, the system momentum located in
the electromagnetic field is introduced by the nonelectromagnetic external forces.

We can also calculate the angular impulse due to the nonelectromagnetic
external forces when the charge $q$ is moved quasistatically in from spatial
infinity. \ The external torque $\Delta\overrightarrow{\Gamma}_{q}^{ext-mag}%
$about the origin due to the force $\Delta\mathbf{F}_{q}^{ext-mag}$ on the
charge $q$\ is given by%
\begin{align}
\Delta\overrightarrow{\Gamma}_{q}^{ext-mag}  &  =\mathbf{r}_{q}\times
\Delta\mathbf{F}_{q}^{ext-mag}=\widehat{i}x_{q}\times\left(  \widehat{j}%
\frac{3q\mu z_{e}^{2}}{c(x_{q}^{2}+z_{e}^{2})^{5/2}}\frac{dx_{q}}{dt}%
-\widehat{j}\frac{q\mu}{c(x_{q}^{2}+z_{e}^{2})^{3/2}}\frac{dx_{q}}{dt}\right)
\nonumber\\
&  =\widehat{k}\frac{q\mu x_{q}}{c}\frac{dx_{q}}{dt}\left(  \frac{3z_{e}^{2}%
}{(x_{q}^{2}+z_{e}^{2})^{5/2}}-\frac{1}{(x_{q}^{2}+z_{e}^{2})^{3/2}}\right)
\end{align}
The external torque on the magnetic moment is due to the nonelectromagnetic
external forces on the charges $e$ and $-\dot{e}.$ \ In the calculation
through the magnetic moment vector $\overrightarrow{\mu}$, the external torque
$\Delta\overrightarrow{\Gamma}_{\mu}^{ext-mag}$ about the origin is due both
to the external force $\Delta\mathbf{F}_{\mu}^{ext-mag}$ acting on
$\overrightarrow{\mu}$ and also to the external torque on $\overrightarrow
{\mu}$ needed to balance the torque $\overrightarrow{\mu}\times\mathbf{B}%
_{q}(\mathbf{r}_{q})$ due to the magnetic field $\mathbf{B}_{q}(\mathbf{r}%
_{q})$ of the moving charge $q.$ \ Thus the average torque about the origin
due to the forces on the charges $e$ and $-e$ is
\begin{align}
\Delta\overrightarrow{\Gamma}_{\mu}^{ext-mag}  &  =\mathbf{r}_{\mu}%
\times\Delta\mathbf{F}_{\mu}^{ext-mag}-\overrightarrow{\mu}\times
\mathbf{B}_{q}(\mathbf{r}_{q})\nonumber\\
&  =\widehat{k}z_{e}\times\left(  -\widehat{j}\frac{q\mu}{c(z_{e}^{2}%
+x_{q}^{2})^{3/2}}\frac{dx_{q}}{dt}\right)  -\mu\widehat{k}\times\left(
q\frac{\mathbf{v}_{q}}{c}\times\frac{-\widehat{i}x_{q}+\widehat{k}z_{e}%
}{(x_{q}{}^{2}+z_{e}^{2})^{3/2}}\right) \nonumber\\
&  =0
\end{align}
Then the total angular impulse $\overrightarrow{\mathcal{I}}^{ext}$ delivered
to the system \ by the nonelectromagnetic forces is that due to the external
force $\Delta\mathbf{F}_{q}^{ext-mag}$ on the charge $q$ alone and is given
by
\begin{align}
\overrightarrow{\mathcal{I}}^{ext}  &  =%
{\displaystyle\int}
dt\Delta\overrightarrow{\Gamma}_{q}^{ext-mag}=%
{\displaystyle\int}
dt\widehat{k}\frac{q\mu x_{q}}{c}\frac{dx_{q}}{dt}\left(  \frac{3z_{e}^{2}%
}{(x_{q}^{2}+z_{e}^{2})^{5/2}}-\frac{1}{(x_{q}^{2}+z_{e}^{2})^{3/2}}\right)
\nonumber\\
&  =%
{\displaystyle\int\limits_{\infty}^{x_{q}}}
dx\widehat{k}\frac{q\mu x}{c}\left(  \frac{3z_{e}^{2}}{(x^{2}+z_{e}^{2}%
)^{5/2}}-\frac{1}{(x^{2}+z_{e}^{2})^{3/2}}\right) \nonumber\\
&  =\widehat{k}\frac{q\mu}{c}\frac{x_{q}^{2}}{(x_{q}^{2}+z_{e}^{2})^{3/2}%
}=\mathbf{r}_{q}\times\left(  \frac{1}{c}\mathbf{E}_{q}\mathbf{(}%
0,0,z_{e})\times\overrightarrow{\mu}\right)
\end{align}
which agrees exactly with the electromagnetic angular momentum calculated in
Eq. (11).

\subsection{Quasistatic Changes for a Point Charge Near a Constant-Current
Solenoid}

It is interesting to note how these results (22)-(29) for a constant-speed
magnetic moment go over to those for an constant-current infinite solenoid.
\ The situation of an infinite solenoid can again be obtained by adding the
contributions of a stack of magnetic moments. \ Thus the force $\Delta
\mathbf{F}_{q}^{ext-sol}$ can be obtained by replacing the magnetic moment
$\mu$ by $dz\,\pi r^{2}B_{0}/(4\pi)$ in Eq. (22) and integrating
\begin{equation}
\Delta\mathbf{F}_{q}^{ext\rightarrow sol}=%
{\displaystyle\int\limits_{-\infty}^{\infty}}
dz\,\left(  \widehat{j}\frac{3qz_{e}^{2}}{c(x_{q}^{2}+z^{2})^{5/2}}%
\frac{dx_{q}}{dt}-\widehat{j}\frac{q}{c(x_{q}^{2}+z^{2})^{3/2}}\frac{dx_{q}%
}{dt}\right)  \frac{\pi r^{2}B_{0}}{4\pi}=0
\end{equation}
Thus in the limit of an infinitely long constant-current solenoid, there is no
external force needed on the charge $q$ since there is no magnetic force on
the charge $q.$ \ The external force $\Delta\mathbf{F}_{\mu}^{ext-sol}$ on the
solenoid needed to balance the magnetic force on the solenoid due to the
moving charge $q$ can be obtained analogously by integrating over the magnetic
moment result in Eq. (24) as%
\begin{align}
\Delta\mathbf{F}_{\mu\rightarrow sol}^{ext}  &  =%
{\displaystyle\int\limits_{-\infty}^{\infty}}
dz(-\widehat{j})\frac{q}{c(z^{2}+x_{q}^{2})^{3/2}}\frac{dx_{q}}{dt}\frac{\pi
r^{2}B_{0}}{4\pi}\nonumber\\
&  =-\widehat{j}\frac{qr^{2}B_{0}}{2cx_{q}^{2}}\frac{dx_{q}}{dt}%
\end{align}
The impulse $\mathbf{I}_{\mu\rightarrow sol}$ delivered to the system by this
nonelectromagnetic external force $\Delta\mathbf{F}_{\mu\rightarrow sol}%
^{ext}$ is%
\begin{equation}
\mathbf{I}_{\mu\rightarrow sol}=%
{\displaystyle\int}
dt\,\Delta\mathbf{F}_{\mu\rightarrow sol}^{ext}=%
{\displaystyle\int\limits_{\infty}^{x_{q}}}
dx\left(  -\widehat{j}\frac{qr^{2}B_{0}}{2cx^{2}}\right)  =\widehat{j}%
\frac{qr^{2}B_{0}}{2cx_{q}}%
\end{equation}
which agrees with the linear momentum in the electromagnetic field calculated
in Eq. (1). \ Thus it is the nonelectromagnetic external force which
introduces the field linear momentum for quasistatic changes in the system of
a point charge and a constant-current solenoid.

The angular impulse $\overrightarrow{\mathcal{I}}_{\mu\rightarrow sol}$
delivered by the nonelectromagnetic external forces can also be calculated.
\ Because of the delicate limit for angular momentum, we give first the result
for a solenoid of finite length $L$ following from Eq. (29)
\begin{align}
\overrightarrow{\mathcal{I}}_{\mu\rightarrow sol}^{(L)}  &  =%
{\displaystyle\int\limits_{-L/2}^{L/2}}
dz\widehat{k}\frac{q}{c}\frac{x_{q}^{2}}{(x_{q}^{2}+z_{e}^{2})^{3/2}}\frac{\pi
r^{2}B_{0}}{4\pi}\nonumber\\
&  =\widehat{k}\frac{qr^{2}B_{0}}{2c}\frac{L/2}{[x_{q}^{2}+(L/2)^{2}]^{3/2}}%
\end{align}
which is in agreement with Eq. (13). \ In the limit of an infinite solenoid,
we have from Eq. (33) that
\begin{equation}
\overrightarrow{\mathcal{I}}_{\mu\rightarrow sol}=\widehat{k}qr^{2}B_{0}/(2c)
\end{equation}
which agrees with the field angular momentum found in Eq. (14).

It is interesting to see the varying roles played by the external forces as
the length of the solenoid increases. \ Thus as the solenoid becomes longer,
the magnetic force on the distant charge $q$ becomes smaller so that all the
electromagnetic field momentum is accounted for by the external force on the
solenoid. \ However, none of the \textit{angular} momentum is introduced into
the system by the external force on the solenoid. \ Rather it is the external
force on the charge $q$ which accounts of the electromagnetic angular
momentum, even in the limit where the solenoid becomes infinite; because of
the extra factor of $x_{q},$ the angular impulse of the torque due to the
external force on $q$ does not vanish even though the linear impulse \ of the
external force on $q$ does indeed vanish.

\section{Varying-Speed Model of a Magnetic Moment}

\subsection{"Hidden Mechanical Momentum" in a Magnetic Moment Model}

In the discussion above, the magnetic moment model involved a charge $e$
moving in its circular orbit with constant speed. \ A variation on this
situation requires that the nonelectromagnetic forces of constraint on the
charge $e$ provide only a centripetal acceleration for the circular orbit.
\ Thus the charge $e$ is allowed to change its speed due to the electric field
of the point charge $q$. \ In this case, the nonelectromagnetic external
forces on the system $\mathbf{F}_{e}^{ext-cent},$ $\mathbf{F}_{-e}^{ext},$ and
$\mathbf{F}_{q}^{ext}$ do no work; this follows since the charge $q$ and the
charge $-e$ are at rest, while the force $\mathbf{F}_{e}^{ext-cent}$ on the
charge $e$ is always perpendicular to the velocity of the charge. \ The
displacement of the charge $e$ is still that of a circular orbit%
\begin{equation}
\mathbf{r}_{e}(t)=r(\widehat{i}\cos\phi+\widehat{j}\sin\phi)+\widehat{k}z_{e}%
\end{equation}
but the velocity must be written as%
\begin{equation}
\mathbf{v}_{e}(t)=r\frac{d\phi}{dt}(-\widehat{i}\sin\phi+\widehat{j}\cos\phi)
\end{equation}
and the acceleration%
\begin{equation}
\mathbf{a}_{e}=-r\left(  \frac{d\phi}{dt}\right)  ^{2}(\widehat{i}\cos
\phi+\widehat{j}\sin\phi)+r\frac{d^{2}\phi}{dt^{2}}(-\widehat{i}\sin
\phi+\widehat{j}\cos\phi)
\end{equation}
Here we are allowing a \textit{response} to the perturbation caused by the
distant charge $q$, and these changes will in general involve changes in
particle position, velocity, and acceleration. \ However, these
\textit{changes} from the unperturbed values will all be first order in the
perturbation when we are carrying out calculations only through first order in
the perturbing field due to the distant charge $q.$ Thus for this new
magnetic-moment model, the electromagnetic energy, linear momentum, and
angular momentum (which were already first order in the perturbation) will
remain unchanged from the values obtained in Part II above. \ What we are now
allowing to change are the mechanical energy, linear momentum and angular
momentum associated with the mass $m$. \ \ 

It is a result from mechanics (to be derived in the next section) that a small
constant perturbing force $\mathbf{F}$ acting on a mass $m$ constrained to
move in a circle leads to a non-zero average (relativistic) mechanical
momentum $\left\langle \mathbf{p}_{m}^{mech}\right\rangle $ for the particle
given by
\begin{equation}
\left\langle \mathbf{p}_{m}^{mech}\right\rangle =\frac{1}{2mc^{2}%
}\mathbf{L\times F}%
\end{equation}
where $\mathbf{L}$ is the average angular momentum of the particle in its
circular motion and the calculation is through first order in the perturbing
force $\mathbf{F}$ and through order $v^{2}/c^{2}$ in the particle speed
$v.$\cite{Gen}\ The angular momentum $\mathbf{L}$ of a particle in orbit is
related to the magnetic moment $\overrightarrow{\mu}$ by\cite{J2}
\begin{equation}
\overrightarrow{\mu}=\frac{e}{2mc}\mathbf{L}%
\end{equation}
so that the mechanical momentum given in Eq. (38) can be rewritten in the case
of a charged particle $e$ in the form
\begin{equation}
\left\langle \mathbf{p}_{m}^{mech}\right\rangle =\frac{1}{2mc^{2}%
}\mathbf{L\times F=}\frac{1}{c}\frac{\overrightarrow{\mu}}{e}\times\mathbf{F}%
\end{equation}
or, in the case that the force $\mathbf{F}$ is provided by a constant electric
field $\mathbf{F=}e\mathbf{E,}$ the relation becomes
\begin{equation}
\left\langle \mathbf{p}_{m}^{mech}\right\rangle \mathbf{=}\frac{1}{c}%
\frac{\overrightarrow{\mu}}{e}\times\mathbf{F=}\frac{1}{c}\overrightarrow{\mu
}\times\mathbf{E}%
\end{equation}
\ In the case of the magnetic moment treated earlier, the electric field
$\mathbf{E}_{q}$ acting on the charge $e$ provides the constant force in the
approximation $r<<x_{q}$ that the perturbing electric field $E_{q}$ due to the
charge $q$ is uniform across the orbital radius $r$ of the charged particle
$e$. \ Thus in the modified magnetic moment model, where the forces of
constraint do no work, there is an average mechanical linear momentum which is
equal in magnitude and opposite in sign to the electromagnetic field momentum
found in Eq. (10). \ For the case of a varying-speed magnetic moment, this
mechanical momentum in Eq. (41) is termed "hidden mechanical momentum" in the
literature. \ In this article, we emphasize that this "hidden mechanical
momentum" is due to the external forces which provide the conditions of
constraint and does not necessarily have anything to do with electromagnetism. \ 

\subsection{"Hidden Mechanical Momentum" for a Constrained Circular Orbit}

We now turn to a derivation of this mechanical momentum of a particle moving
along a prescribed path under an external perturbing force. \ In order to
emphasize that this mechanical momentum need not be related to
electromagnetism, we will phrase our discussion of the particle motion in
terms of a general perturbing force $\mathbf{F}$ rather than in terms of a
force $e\mathbf{E}$ due to an electric field $\mathbf{E}$. \ We consider a
particle of mass $m$ constrained by external centripetal forces to move in a
circle of radius $r$ parallel to the $xy$-plane while subjected to a small
constant perturbing force $\mathbf{F=-}\widehat{\mathbf{i}}f+\widehat{j}%
F_{z}\mathbf{~}$which has a component $-f$ parallel to the $x$-axis. \ The
expressions for the displacement, velocity, and acceleration of the mass $m$
are take exactly as given above in Eqs. (35)-(37), except that we drop the
subscript $e$.

The speed of the mass $m$ follows from energy conservation as
\begin{equation}
m\gamma_{0}c^{2}=m\gamma c^{2}+rf\cos(\phi)
\end{equation}
where $\gamma=[1-(v^{2}/c^{2})]^{-1/2.}$ Then writing $v=v_{0}+\Delta v$ and
retaining terms through first order in $v_{0}^{2}/c^{2}$ and first order in
$rf/(mv_{0}^{2}),$ we may replace $\phi$ in Eq. (42) by the unperturbed value
$\omega_{0}t$ and obtain
\begin{equation}
v=r\frac{d\phi}{dt}=v_{0}-\frac{rf}{mv_{0}}\left(  1-\frac{3}{2}\frac
{v_{0}^{2}}{c^{2}}\right)  \cos(\omega_{0}t)
\end{equation}
We can integrate equation (43) once with respect to time to obtain
$\phi=\omega_{0}t+\Delta\phi,$ given through first order in the perturbation
$f$ as
\begin{equation}
\phi(t)=\omega_{0}t-\frac{f}{mv_{0}\omega_{0}}\left(  1-\frac{3}{2}\frac
{v_{0}^{2}}{c^{2}}\right)  \sin(\omega_{0}t)
\end{equation}
The direction $(-\widehat{i}\sin\phi+\widehat{j}\cos\phi)$ of the tangential
velocity $\mathbf{v}$\ can be evaluated from Eq. (44) for $\phi$ and the use
of the first order expansions
\begin{equation}
\sin(\phi+\Delta\phi)=\sin(\phi)+\Delta\phi\cos(\phi)
\end{equation}
and
\begin{equation}
\cos(\phi+\Delta\phi)=\cos(\phi)-\Delta\phi\sin(\phi)
\end{equation}
giving%
\begin{align}
\lbrack-\widehat{i}\sin\phi+\widehat{j}\cos\phi]  &  =-\widehat{i}\sin
(\omega_{0}t)+\widehat{j}\cos(\omega_{0}t)\nonumber\\
&  +\frac{f}{mv_{0}\omega_{0}}\left(  1-\frac{3}{2}\frac{v_{0}^{2}}{c^{2}%
}\right)  \sin(\omega_{0}t)[\widehat{i}\cos(\omega_{0}t)+\widehat{j}%
\sin(\omega_{0}t)
\end{align}

We can now calculate the average mechanical linear momentum of the mass $m$
which is moving in a circular orbit. \ Through first order in the perturbing
force $f$ and first order in $v_{0}^{2}/c^{2},$ the mechanical momentum
$\mathbf{p}_{m}^{mech}$ is given from Eqs. (36), (42), and (47) by
\begin{align}
\mathbf{p}_{m}^{mech}  &  =m\gamma_{e}\mathbf{v}_{e}\nonumber\\
&  =\left(  m\gamma_{0}-\frac{rf}{c^{2}}\cos(\omega_{0}t\right)  \left(
v_{0}-\frac{rf}{mv_{0}}\left(  1-\frac{3}{2}\frac{v_{0}^{2}}{c^{2}}\right)
\cos(\omega_{0}t)\right)  [-\widehat{i}\sin\phi+\widehat{j}\cos\phi
]\nonumber\\
&  =\left(  m\gamma_{0}v_{0}-\frac{v_{0}rf}{c^{2}}\cos(\omega_{0}%
t)-\frac{\gamma_{0}rf}{v_{0}}\left(  1-\frac{3}{2}\frac{v_{0}^{2}}{c^{2}%
}\right)  \cos(\omega_{0}t)\right)  [-\widehat{i}\sin(\omega_{0}t)+\widehat
{j}\cos(\omega_{0}t)\nonumber\\
&  +m\gamma_{0}v_{0}\frac{f}{mv_{0}\omega_{0}}\left(  1-\frac{3}{2}\frac
{v_{0}^{2}}{c^{2}}\right)  \sin(\omega_{0}t)[\widehat{i}\cos(\omega
_{0}t)+\widehat{j}\sin(\omega_{0}t)
\end{align}
Then averaging in time gives us an average mechanical momentum for the mass
$m$%
\begin{equation}
\left\langle \mathbf{p}_{m}^{mech}\right\rangle =-\widehat{j}\frac{v_{0}%
rf}{2c^{2}}=-\widehat{j}\frac{Lf}{2mc^{2}}=\frac{1}{2mc^{2}}\mathbf{L\times F}%
\end{equation}
where $\mathbf{L}=\widehat{k}rmv_{0}$ and $\mathbf{F=-}\widehat{\mathbf{i}%
}f+\widehat{j}F_{z}$ in agreement with Eq. (38).

We notice that there is no change in the average angular momentum due to the
perturbing force $\mathbf{F}$. \ Thus the average angular momentum is
\begin{align}
\left\langle \mathbf{L}\right\rangle  &  =\left\langle \widehat{k}rm\gamma
v\right\rangle =\widehat{k}rm\left\langle v+\frac{v^{3}}{2c^{2}}\right\rangle
\nonumber\\
&  =\widehat{k}rm\left\langle v_{0}+\frac{v_{0}^{3}}{2c^{2}}+\Delta v\left(
1+\frac{3}{2}\frac{v_{0}^{2}}{c^{2}}\right)  \right\rangle =\mathbf{L}_{0}%
\end{align}
since from Eq. (43) $\left\langle \Delta v\right\rangle =0.$

\subsection{External Forces for the "Hidden Mechanical Momentum" \ }

The mechanical system discussed in the previous section consists of one point
mass $m$ in a circular orbit, and the energy, linear momentum, and angular
momentum are all associated with this point mass alone. \ Thus as far as our
mechanical system is concerned, \textit{all} forces on the mass $m$ are
external forces, both the constant perturbing force $\mathbf{F=-}\widehat
{i}f+\widehat{j}F_{z}$ and the centripetal constraining force $\mathbf{F}%
^{cent}.$ The time-average values $\left\langle \mathbf{F\cdot v}\right\rangle
,$ $\left\langle \mathbf{r\times F}\right\rangle ,$ and $\left\langle
(\mathbf{F\cdot v})\mathbf{r}\right\rangle $ can all be calculated from the
results of Eqs. (43) - (47). \ By symmetry alone, we can easily see that
$\left\langle \mathbf{F\cdot v}\right\rangle =0,$ and $\left\langle
\mathbf{r\times F}\right\rangle =0.$ \ However, from Eqs. (43) - (47), we see
that through first order in the perturbation $f$
\begin{align}
\mathbf{F\cdot v}  &  =(\mathbf{-}\widehat{i}f+\widehat{j}F_{z})\cdot
\{[-\widehat{i}\sin(\omega_{0}t)+\widehat{j}\cos(\omega_{0}t)]\left(
v_{0}-\frac{rf}{mv_{0}}\left(  1-\frac{3}{2}\frac{v_{0}^{2}}{c^{2}}\right)
\cos(\omega_{0}t)\right) \nonumber\\
&  +\frac{f}{mv_{0}\omega_{0}}\left(  1-\frac{3}{2}\frac{v_{0}^{2}}{c^{2}%
}\right)  \sin(\omega_{0}t)[\widehat{i}\cos(\omega_{0}t)+\widehat{j}%
\sin(\omega_{0}t)]v_{0}\}\nonumber\\
&  =-fv_{0}\sin(\omega_{0}t)
\end{align}
and so
\begin{align}
\left\langle (\mathbf{F\cdot v)r}\right\rangle  &  =_{av}/[fv_{0}\sin
(\omega_{0}t)]r\{[\widehat{i}\cos(\omega_{0}t)+\widehat{j}\sin(\omega
_{0}t)]+\widehat{k}z_{e}\}\backslash_{av}\nonumber\\
&  =\widehat{j}\frac{rfv_{0}}{2}%
\end{align}
The centripetal constraining force $\mathbf{F}^{cent}$ can be found from the
equation of motion for the mass $m$, $(d/dt)(m\gamma\mathbf{v)=F}%
^{cent}+\mathbf{F}$, giving%
\begin{equation}
\mathbf{F}^{cent}=\frac{d}{dt}(m\gamma\mathbf{v)}-\mathbf{F}%
\end{equation}
Since the particle motion is periodic in time and the perturbing force
$\mathbf{F}$ is constant, it follows from Eq. (53) that on time average
$\mathbf{F}^{cent}$ balances the perturbing force, $\left\langle
\mathbf{F}^{cent}\right\rangle =\mathbf{F}.$ \ Also, since $\mathbf{F}^{cent}$
is a centripetal force, we find vanishing values for\textbf{ }$\mathbf{F}%
^{cent}\cdot\mathbf{v=}0$, $\mathbf{r\times F}^{cent}=0,$ and ($\mathbf{F}%
^{cent}\cdot\mathbf{v)r}=0.$ \ These results allow us to confirm (on time
average) the conservation laws for energy, linear momentum, and angular
momentum for the system consisting of the mass $m$. \ The one interesting
result follows from relativistic symmetry corresponding to Eq. (4). \ In the
present case, there is no average change in energy or center of energy so that
the time average of Eq. (4) involves the average power-weighted displacement
given by Eq. (52) balancing the average linear momentum in Eq. (49)%
\begin{equation}
\left\langle (\mathbf{F\cdot v)r}\right\rangle =\widehat{j}\frac{rfv_{0}}%
{2}=-c^{2}\left\langle \mathbf{p}_{m}^{mech}\right\rangle
\end{equation}

\subsection{Quasistatic Increase of the Perturbing Force}

It seems somewhat surprising to find that a particle which is required by
centripetal forces of constraint to remain in a circular orbit despite the
influence of a small constant perturbing force should acquire an average
linear momentum in the plane of the orbit in a direction perpendicular to the
perturbing force, corresponding to Eq. (49). \ In order to confirm our result,
we will consider the average impulse introduced by the external forces of
constraint when the perturbing force $\mathbf{F=-}\widehat{i}f+\widehat
{j}F_{z}$ is increased quasistatically from zero. \ Thus we will write the
changing $x$-component of the perturbing force as $\alpha t$ where $\alpha$ is
taken as very small and the total time $t_{f}$ of increase is large so that
the final value is $f=\alpha t_{f}.$

The equation of motion of the mass $m$ is $(d/dt)(m\gamma\mathbf{v)=F}%
^{cent}+\mathbf{F}$ where $\mathbf{F}^{cent}$ is the centripetal force of
constraint which keeps the mass $m$ in a circular orbit but is always
perpendicular to the orbit, and $\mathbf{F}=\mathbf{-}\widehat{i}\alpha
t+\widehat{j}F_{z}$ is the slowly-changing, spatially uniform perturbing force
on the orbit. \ The point mass $m$ remains in a plane parallel to the
$xy$-plane, and therefore the $z$-component of $\mathbf{F}^{cent}$ simply
balances the $z$-component of $\mathbf{F,}$ $F_{z}^{cent}=-F_{z}.$ \ In the
plane of the particle motion, the equation of motion becomes%
\begin{equation}
\frac{d}{dt}(m\gamma\mathbf{v)=-}(\widehat{i}\cos\phi+\widehat{j}\sin
\phi)F_{xy}^{cent}-\widehat{i}\alpha t
\end{equation}
If we take the inner product of this equation with the particle velocity
$\mathbf{v}$, then we obtain the energy equation%
\begin{equation}
\mathbf{v\cdot}\frac{d}{dt}(m\gamma\mathbf{v)=}\frac{d}{dt}(m\gamma
c^{2})\mathbf{=v}\cdot(-\widehat{i}\sin\phi+\widehat{j}\cos\phi)F_{xy}%
^{cent}-\mathbf{v}\cdot\widehat{i}\alpha t=v\alpha t\sin\phi
\end{equation}
since $\mathbf{F}^{cent}$ is always perpendicular to the particle velocity.
\ \ Since the orbital radius $r$ is not changing, this equation can be
rewritten through order $v^{2}/c^{2}$ as%
\begin{equation}
mv\left(  1+\frac{3v^{2}}{2c^{2}}\right)  r\frac{d^{2}\phi}{dt^{2}}=v\alpha
t\sin\phi
\end{equation}
or%
\begin{equation}
\frac{d^{2}\phi}{dt^{2}}=\left(  1-\frac{3v^{2}}{2c^{2}}\right)  \frac{\alpha
t}{mr}\sin\phi
\end{equation}
Since the right-hand side of Eq. (58) is already first order in the
perturbation $\alpha,$ we may replace $\phi$ on the right-hand side by
$\phi=\omega_{0}t$ and $v$ by $v=v_{0}.$ \ We can then integrate to obtain the
angular velocity%
\begin{equation}
\frac{d\phi}{dt}=\frac{v}{r}=\omega_{0}+\left(  1-\frac{3v_{0}^{2}}{2c^{2}%
}\right)  \frac{\alpha}{m\omega_{0}r}\left(  \frac{\sin(\omega_{0}t)}%
{\omega_{0}^{2}}-\frac{t\cos\left(  \omega_{0}t\right)  }{\omega_{0}}\right)
\end{equation}
and the angle $\phi$
\begin{equation}
\phi=\omega_{0}t+\left(  1-\frac{3v_{0}^{2}}{2c^{2}}\right)  \frac{\alpha}%
{mr}\left(  -\frac{2\cos(\omega_{0}t)}{\omega_{0}^{3}}-\frac{t\sin(\omega
_{0}t)}{\omega_{0}^{2}}\right)
\end{equation}
\ We notice that when we take the time $t=t_{f}$ with the limit of $\alpha$
vanishingly small but $f=\alpha t_{f},$ we obtain the results in Eqs. (43) and (44).

The force $\mathbf{F}^{cent}$ provides the centripetal acceleration following
from Eq. (55) as%
\begin{align}
\mathbf{F}_{xy}^{cent}  &  =-(\widehat{i}\cos\phi+\widehat{j}\sin\phi
)m\gamma\frac{v^{2}}{r}+(\widehat{i}\cos\phi+\widehat{j}\sin\phi)\alpha
t\cos\phi\nonumber\\
&  =-(\widehat{i}\cos\phi+\widehat{j}\sin\phi)m\left(  1+\frac{v^{2}}{2c^{2}%
}\right)  \frac{v^{2}}{r}+(\widehat{i}\cos\phi+\widehat{j}\sin\phi)\alpha
t\cos\phi
\end{align}
Once again we need to use the first order expansions for the sine and cosine
of $\phi=\phi+\Delta\phi$ given in Eqs. (45) and (46), and the first order
expansions of powers of $v=v_{0}+\Delta v$%
\begin{align}
\mathbf{F}_{xy}^{cent}  &  =-\{\widehat{i}\cos(\omega_{0}t)+\widehat{j}%
\sin(\omega_{0}t)\}\frac{m}{r}\left[  v_{0}^{2}\left(  1+\frac{v_{0}^{2}%
}{2c^{2}}\right)  +2v_{0}\Delta v\left(  1+\frac{v_{0}^{2}}{c^{2}}\right)
\right] \nonumber\\
&  +\{\widehat{i}\sin(\omega_{0}t)+\widehat{j}\cos(\omega_{0}t)\}\Delta
\phi\frac{mv_{0}^{2}}{r}\left(  1+\frac{v_{0}^{2}}{2c^{2}}\right)
+[\widehat{i}\cos(\omega_{0}t)+\widehat{j}\sin(\omega_{0}t)]\alpha t\cos
\phi\nonumber\\
&  =-\{\widehat{i}\cos(\omega_{0}t)+\widehat{j}\sin(\omega_{0}t)\}\frac{m}%
{r}\left[  v_{0}^{2}\left(  1+\frac{v_{0}^{2}}{2c^{2}}\right)  \right]
+[\widehat{i}\cos(\omega_{0}t)+\widehat{j}\sin(\omega_{0}t)]\alpha t\cos
\phi\nonumber\\
&  --\{\widehat{i}\cos(\omega_{0}t)+\widehat{j}\sin(\omega_{0}t)\}\frac{m}%
{r}2v_{0}^{3}\left(  1+\frac{v_{0}^{2}}{c^{2}}\right)  \left(  1-\frac
{3v_{0}^{2}}{2c^{2}}\right)  \frac{\alpha}{m\omega_{0}}\left(  \frac
{\sin(\omega_{0}t)}{\omega_{0}^{2}}-\frac{t\cos\left(  \omega_{0}t\right)
}{\omega_{0}}\right) \nonumber\\
&  +\{\widehat{i}\sin(\omega_{0}t)+\widehat{j}\cos(\omega_{0}t)\}\frac
{mv_{0}^{2}}{r}\left(  1+\frac{v_{0}^{2}}{2c^{2}}\right)  \left(
1-\frac{3v_{0}^{2}}{2c^{2}}\right)  \frac{\alpha}{mr}\left(  -\frac
{2\cos(\omega_{0}t)}{\omega_{0}^{3}}-\frac{t\sin(\omega_{0}t)}{\omega_{0}^{2}%
}\right)
\end{align}
If we now average over time, we find that Eq. (62) becomes%
\begin{equation}
\left\langle \mathbf{F}_{xy}^{cent}\right\rangle =\widehat{i}\alpha
t-\widehat{j}\frac{rv_{0}}{2c^{2}}\alpha=\widehat{i}f-\widehat{j}\frac{rv_{0}%
}{2c^{2}}\frac{df}{dt}%
\end{equation}
We see from Eq. (63) that the average external force on the mass $m$ balances
the perturbing force $\mathbf{F}$ \ and also has a second contribution
proportional to the rate of change of the perturbing force $\mathbf{F}$%
\begin{equation}
\left\langle \mathbf{F}^{cent}\right\rangle =-\mathbf{F+}\frac{1}{2mc^{2}%
}\mathbf{L\times}\frac{d\mathbf{F}}{dt}%
\end{equation}
\ If the perturbing force $\mathbf{F}$ is increased quasistatically, then (on
time average) the constraint-maintaining force $\mathbf{F}^{cent}$ balances
the perturbing force $\mathbf{F}$ and also introduces a net unbalanced impulse
$\mathbf{I}^{cent-\mathbf{F}}$%
\begin{align}
\mathbf{I}^{cent-\mathbf{F}}  &  =%
{\displaystyle\int}
dt\left\langle \mathbf{F}^{cent}\right\rangle =%
{\displaystyle\int}
dt\frac{1}{2mc^{2}}\mathbf{L\times}\frac{d\mathbf{F}}{dt}\,\nonumber\\
&  =\frac{1}{2mc^{2}}\mathbf{L\times F}%
\end{align}
which exactly accounts for the average linear momentum stored in the
constrained motion as given in Eq. (49). \ It is emphasized that this
mechanical momentum is introduced by the external force of constraint
$\mathbf{F}^{cent}$ which keeps the charge $e$ moving precisely in a circular
orbit, and not by the perturbing force $\mathbf{F}$ which is in a direction
perpendicular to the average momentum. \ On the other hand, the power
delivered by the perturbing force $\mathbf{F}$ is associated with the average
linear momentum as in Eq. (52).

\subsection{Electric Dipole Moment Produced by External Forces of Constraint
in the "Hidden Momentum" Case}

In addition to producing a "hidden momentum," the nonelectromagnetic external
forces of constraint in our varying-speed model for a magnetic moment also
produce an average electric dipole moment which seems to go unmentioned in the
literature. \ Thus if a current is constant but the speed of the charges is
varying in space, then the charge density must also vary in space. \ For our
example involving circular motion, the time-average linear charge density
$\lambda(\phi)$ on the circular path followed by the charge $e$ is inversely
related to the speed of the charge at the angle $\phi.$ \ Then from Eq. (44),
with the reinsertion of $f=eE_{q},$ we have to first order in the perturbing
field $\mathbf{E}_{q}$%
\begin{equation}
\lambda(\phi)=\frac{e}{2\pi r}\frac{\left\langle v\right\rangle }{v}=\frac
{e}{2\pi r}\left[  1+\frac{reE_{q}}{mv_{0}^{2}}\left(  1-\frac{3}{2}%
\frac{v_{0}^{2}}{c^{2}}\right)  \cos\phi\right]
\end{equation}
Then the time-average electric dipole moment for our varying-speed magnetic
moment is%
\begin{align}
\left\langle \overrightarrow{\mathfrak{p}}\right\rangle  &  =%
{\displaystyle\int}
(d\phi\,r)\mathbf{r}\lambda(\phi)\nonumber\\
&  =%
{\displaystyle\int}
(d\phi\,r)(\widehat{i}\cos\phi+\widehat{j}\sin\phi)\frac{e}{2\pi r}\left[
1+\frac{reE_{q}}{mv_{0}^{2}}\left(  1-\frac{3}{2}\frac{v_{0}^{2}}{c^{2}%
}\right)  \cos\phi\right] \nonumber\\
&  =\widehat{i}\frac{e^{2}rE_{q}}{2mv_{0}^{2}}\left(  1-\frac{3}{2}\frac
{v_{0}^{2}}{c^{2}}\right)
\end{align}
Alternatively, we can calculate the electric dipole moment in Eq. (67) by
taking the time average of $e\mathbf{r}_{e}=e(\widehat{i}\cos\phi+\widehat
{j}\sin\phi)=e(\widehat{i}\cos\omega_{0}t+\widehat{j}\sin\omega_{0}%
t)+e\Delta\phi(-\widehat{i}\cos\omega_{0}t+\widehat{j}\sin\omega_{0}t),$ where
$\Delta\phi=\phi-\omega_{0}t$ is read off from Eq. (44). \ We note that this
average electric dipole moment involves nonrelativistic terms as well as terms
in order $1/c^{2}.$ \ The nonrelativistic terms give an electric dipole moment
in the \textit{opposite} direction from that expected when discussing the
polarization of a conductor due to an external charge $q.$ \ The expression
diverges as the speed of the charge $e$ decrease toward zero; however, this
zero-speed \ limit is inconsistent with our assumption that we are expanding
in a small perturbation where $erE_{q}/(mv_{0}^{2})<<1.$ \ This electric
dipole will put a force on the distant charge $q$ which is causing the
polarization of the magnetic moment.

It seems worth noting that any proposal for "hidden mechanical momentum" which
depends upon the electrical current density $\mathbf{J}=ne\mathbf{v}$
remaining constant while the speed $v$ varies along a prescribed path requires
that the density $n$ of charges $e$ must vary along the path and so must
produce an electric dipole moment connected to the perturbing electric field
and dependent upon the unperturbed speed $v_{0}$ analogous to the result in
Eq. (67). \ It is also worth emphasizing that the direction of this magnetic
moment depends explicitly upon the presence of external centripetal forces of
constraint. \ In the case of a purely electromagnetic magnetic moment based
upon a classical hydrogen atom, the magnetic moment also develops an electric
dipole moment due to a perturbing electric field, but this dipole moment is in
a direction \textit{perpendicular} (not parallel) to the perturbing
field.\cite{S}

\subsection{Momentum Associated with Work Done by External Forces in Magnetic
Moment Models}

In the models which we have considered, the existence of linear momentum is
associated with work done by external forces according to the law given in Eq.
(4). \ In the example of Part II where the particle is moving with constant
speed in a circular orbit, the work is done by the external forces
$\mathbf{F}_{e}^{ext}$ which maintain the constant speed of the particle. \ In
this case, the system consists of both electromagnetic fields and mechanical
masses so that the electric field is internal to the system. \ In this case,
the work of the electric field is done against the external force of
constraint which removes energy from the system when the particle $e$\ has
positive values of coordinate $y$ and introduces energy into the system when
the particle $e$ has negative values of coordinate $y.$ \ Thus the value of $%
{\displaystyle\sum}
(\mathbf{F}_{i}^{ext}\cdot\mathbf{v}_{i})\mathbf{r}_{i}$ is in the negative
$y$-direction as in Eq. (19). \ On the other hand, in the example for the
"hidden mechanical momentum" in Parts IIB and IIC where the particle $m$ moved
in a circle with varying speed, the system consists of the mass $m$ alone so
that both the perturbing force $\mathbf{F}$ and the centripetal forces of
constraint are external to the system. \ In this case, it is the perturbing
force $\mathbf{F}$ which is an external force doing work on the system,
introducing kinetic energy when the particle has positive values of coordinate
$y$ and removing kinetic energy when the particle has negative values of
coordinate $y.$ \ Thus the value of $%
{\displaystyle\sum}
(\mathbf{F}_{i}^{ext}\cdot\mathbf{v}_{i})\mathbf{r}_{i}=\left\langle
(\mathbf{F\cdot v})\mathbf{r}\right\rangle $ is in the positive $y$-direction
in this second case as in Eq. (52)

Although both the field momentum of Eq. (10) and the "hidden mechanical
momentum" of Eq. (41) are associated with the work done by external forces, it
should be emphasized that the "hidden mechanical momentum" actually arises
from a different level of approximation than the electromagnetic field
momentum which appeared above. \ The electromagnetic field momentum due to the
charge $q$ and the magnetic moment arises in a calculation based upon the
\textit{unperturbed} motion of the magnetic moment and the point charge, and
therefore is present in the magnetic moment examples of both Parts II and
III\ discussed above. \ This is not the case with the "hidden mechanical
momentum." \ For the "hidden mechanical momentum," the motion of the mass $m$
must \textit{respond} to the perturbing force before the new mechanical linear
momentum appears. \ Thus in the earlier discussions where the magnetic moment
was produced by a charge undergoing uniform circular motion, there is no such
"hidden momentum" at all. \ The distinction which is being made here is by no
means trivial. \ In the case of a magnetic moment modeled as a hydrogen
atom,\cite{B-JP} the electromagnetic field momentum indeed arises because it
appears from the unperturbed motion; however, the "hidden mechanical momentum"
is not found because the nonrelativistic perturbed motion is quite different
from that of a rigid circular orbit.

\section{Discussion}

\subsection{Overview of Charge-Magnet Interactions}

At this point we have seem to have reached the end of our careful analysis for
the two simple models for magnetic systems, and it is appropriate to place the
analysis in perspective. \ The sharp increase in interest\cite{R} regarding
the interaction of charges and magnets goes back at least to the 1960s, and
particularly to the paradox emphasized by Shockley and James\cite{SJ} in their
article entitled, "'Try simplest cases' discovery of 'hidden momentum' forces
on 'magnetic currents.'" \ In an oft-cited response, Coleman and Van
Vleck\cite{CV} point out that the electromagnetic momentum is of order
$1/c^{2},$ and therefore the mechanical behavior of the system must be treated
relativistically to the same order. \ These authors show that the conservation
law for linear momentum need not be violated, without providing a detailed
account for the behavior of the charge-magnet system. \ They mention in a foot
note exactly the varying-speed model for a magnetic moment which is discussed
in the present article in Part III (but they make no reference to the forces
of constraint). \ Furry's extensive analysis followed shortly
thereafter.\cite{F}

Discussions of the interaction of a charge and a magnet usually assume that
the system is \textit{stationary} and \textit{closed} (having \textit{no}
\textit{external} forces of constraint). \ These two assumption are unlikely
to be compatible physically. The charge-magnet system may well be unstable,
just as a point charge outside a conductor is unstable. \ However, these
treatments (which assume closed, stationary behavior) note that the total
system momentum must vanish; therefore there must be momentum present to
balance any arising from the electric field of the distant charge $q$ and the
magnetic field of the magnet.\cite{AR} The theorem does not specify the nature
of the momentum. \ It is interesting that Johnson, Cragin, and
Hodges\cite{JCH} avoid reference to "hidden mechanical momentum." \ Rather,
they repeatedly follow the flow of electromagnetic energy within the systems
which they discuss. \ Furry\cite{F} does refer to "material momentum" and
"hidden momentum" but never insists that the momentum associated with the
energy flow is "mechanical" momentum.

In any case, the "'hidden momentum' forces" mentioned in the title of Shockley
and James's article have now become the "hidden mechanical momentum" described
in the most recent editions of electromagnetism textbooks.\cite{G}\cite{J}
\ It seems curious that this concept of "hidden mechanical momentum" does not
seem to appear in any mechanics textbook but rather only in electromagnetism
texts. \ The reason for this may be conjectured. \ Although the "hidden
mechanical momentum" appearing in Eqs. (38), (41) and (49) may be an
appropriate subject for mechanics, the ideas and dependence upon external
forces of constraint (constraints accurate to order $1/c^{2})$ seem quite
contrived for a realistic discussion of mechanical forces and special
relativistic effects. \ However, electromagnetism incorporates relativity as
soon as one passes beyond electrostatics. \ Furthermore, electromagnetism
often calculates electromagnetic fields from prescribed charge and current
distributions without inquiring about the forces of constraint necessary to
produce the prescribed distributions. \ Finally, electromagnetism is far more
poorly understood than mechanics, and therefore contrived apparent solutions
may be more tolerable when facing unsolved problems. \ The idea of "hidden
mechanical momentum" has passed into the electromagnetism literature in its
present form because of our inability to describe in detail the interaction of
a point charge and a solenoid. \ Furthermore, there are now physicists who
have strong vested interests in certain points of view regarding these interactions.

\subsection{The Accepted \ View as Related to Controversy Over the
Aharonov-Casher Phase Shift}

The motivating force for the presently accepted description of charge-magnet
interactions comes from controversy over the Aharonov-Casher effect\cite{AC}
suggested in 1984 and observed experimentally\cite{Werner} in 1989. \ This
effect is claimed\cite{AR} to be the dual of the Aharonov-Bohm effect\cite{AB}
suggested in 1959 and first observed experimentally\cite{C} in 1960. \ Both
experiments involve the interactions of charges and magnets \ Although both
observed experimental\ effects involve the shift of a particle interference
pattern, the theoretical basis for the shifts is in dispute. \ On one side of
the dispute are those who claim that these phase shifts represent quantum
topological effects with no classical analogue.\cite{AR2} \ On the other side
are those who claim that the phase shifts may well arise from the classical
electromagnetic interactions of charges and magnets.\cite{B-JP} \ This
controversy had existed earlier in connection with the Aharonov-Bohm
effect,\cite{B1973b} but became acute in 1987 when it was pointed
out\cite{B1987} that the calculation of Aharonov and Casher assumed a magnetic
moment made of magnetic charges; for a current-loop magnetic moment, naive
ideas of classical electromagnetism and Newtonian mechanics would account for
the phase shift of the Aharonov-Casher effect.\cite{ARerror} \ In order to
counter this classical electromagnetic argument, proponents of the quantum
\ topological view suggested\cite{APV} that "hidden mechanical momentum"
rendered the naive, current-loop classical analysis invalid. \ In the late
1980s both sides of the controversy submitted articles to both the Physical
Review and to the American Journal of Physics. \ It seems curious that at that
time The Physical Review accepted the quantum article and rejected the
classical article whereas the American Journal of Physics did just the
opposite.\cite{Cur} \ However, in 1990, the American Journal of Physics
accepted the "hidden mechanical momentum" analysis\cite{V} of the
charge-magnet interaction which was favored by the quantum side of the
dispute, and this point of view now dominates all the influential literature
and textbooks to the exclusion of the classical point of view.\cite{FP}

The presently accepted view is that there is no force between a charge and a
long magnet, and that "hidden mechanical momentum" plays a significant role in
realizing this situation.\cite{AR3} \ Within the accepted view, the crucial
role played by "hidden mechanical momentum" can be understood as follows.
\ When a point charge and a long but finite-length magnet are far away, there
is no linear momentum of interaction between them. \ However, if the charge is
in motion toward the magnet, then, when the charge approaches the magnet, (due
to the overlap of the electric field of the charge and the magnetic field of
the magnet) there will be electromagnetic linear momentum in the system. \ The
change in this field linear momentum suggests that there must have been an
interaction between the charge and the magnet which might be compensated by
motions of the centers of energy of the charge and/or of the magnet. \ Indeed,
the magnetic field of the moving charge exerts an obvious Lorentz force on the
magnet. \ Now such an interaction violates the view that there is no force
(causing a change in motion) between a charge and a long magnet. \ Thus here
is where "hidden mechanical momentum" comes in. \ The presently accepted view
claims that the (current-loop) magnet acquires a "hidden mechanical momentum"
which exactly "compensates" the electromagnetic field momentum associated with
the Lorentz force on the magnet. \ In other words, the momentum transferred to
the magnet by the magnetic Lorentz force is in the form of "hidden mechanical
momentum" which involves no motion of the magnet's center of energy.
\ Therefore the total momentum of the system remains zero, consistent with the
original Aharonov-Casher view that there is no force (causing a change in
motion) between the charge and the magnet. \ This view is contained in the
claim\cite{APV}\cite{V}\cite{H}\cite{AR} that the "force" on a magnetic moment
$\overrightarrow{\mu}$ due to a magnetic field is not just the Lorentz force
$\nabla(\overrightarrow{\mu}\cdot\mathbf{B})$ as reported in the previous
editions of the electromagnetism textbooks but rather is $\nabla
(\overrightarrow{\mu}\cdot\mathbf{B})+(1/c^{2})(d/dt)(\mathbf{E\times
}\overrightarrow{\mathbf{\mu}}\mathbf{)}$ so that for a constant magnetic
moment the "force" on a magnetic moment becomes ($\overrightarrow{\mu}%
\cdot\nabla)\mathbf{B,}$ that of a magnetic moment formed from magnetic
charges. \ 

The present article casts a skeptical eye on the accepted point of view that
there are no forces exchanged between charges and long magnets, and that the
force on a magnetic moment is that of the magnetic-charge model. \ It
encourages skepticism for four basic reasons. \ 1) In the analysis above where
we looked at two standard models for a (current-loop) magnetic moment, we have
found that these models require the presence of external forces of constraint,
and that it is precisely these external forces which introduce the "hidden
mechanical momentum" into the charge-magnet system. \ How can we be sure that
a magnet appearing in nature will actually have nonelectromagnetic forces
which \textit{respond} in precisely the fashion demanded by the presently
accepted view? \ It should be\ emphasized that although electromagnetic linear
momentum arises from the unperturbed behavior of a charge and magnet, the
"hidden mechanical momentum" arises from a \textit{perturbation} of the
magnetic moment behavior, and this perturbation must take a rigidly prescribed
form. \ 2) The accepted view involving "hidden mechanical momentum" is
certainly in error because it does not properly account for the angular
momentum in the electromagnetic field for a point charge which approaches a
long but finite-length magnet from spatial infinity. \ Thus in Section IIG we
found that upon quasistatic motion of the charge $q$, the external forces of
constraint (which balanced the magnetic forces between the charge and magnet)
introduced electromagnetic field angular momentum into the charge-magnet
system. \ This electromagnetic field angular momentum is based upon the
unperturbed behavior of the charge and of the magnetic moment, and therefore
is, in lowest order of approximation, also present even when the external
forces are absent. \ However, the external forces of constraint in Section
IIIB which introduce the "hidden mechanical momentum" do not introduce any
angular momentum. \ Thus the contrived argument about "hidden mechanical
momentum" balancing the electromagnetic field linear momentum may allow the
total linear momentum to remain zero as the charge approaches the magnet, but
it can not serve this function for angular momentum which vanished when the
charge was very far away from a (finite-length) magnet. \ Thus there is (at
the very least) a missing link in the presently-accepted statements that there
is no interaction between a charge and a long magnet, and that the force on a
magnetic moment is properly given as $\nabla(\overrightarrow{\mu}%
\cdot\mathbf{B})+(1/c^{2})(d/dt)(\mathbf{E\times}\overrightarrow{\mathbf{\mu}%
}\mathbf{).}$ \ 3) The accepted view prescribes a \textit{perturbed} motion
for the magnetic moment which unavoidably produces an electric dipole moment.
\ This electric dipole moment will apply a (zero-order in $1/c^{2})$
current-dependent electrostatic force back on the distant charge $q$, in
contradiction to the claim that there is no force between a charge and a long
magnet. \ 4) The presently accepted view of charge-magnet interactions demands
a nonrelativistic behavior which is wildly different from that found in the
simplest model of a magnetic moment. \ The simplest model of a magnetic moment
has no external forces and corresponds to a classical hydrogen atom. \ This
version of a magnetic moment can be obtained from the models considered above
by choosing the central charge $-e$ to have large mass compared to the mass
$m$ of the orbiting charge $e$ and by choosing the speed of the charge $e$ so
that $m\gamma v^{2}/r=e^{2}/r^{2}$, placing the light mass (charge $e$) in
Coulomb orbit about the heavy mass (charge $-e$). \ The behavior of this
purely electromagnetic magnetic moment has been analyzed by Solem\cite{S} and
is found to have a nonrelativistic behavior totally different from that
demanded by the promoters of "hidden momentum." \ The interaction between this
magnetic moment and a point charge has been calculated in some
detail,\cite{B-JP} and there is indeed an electric force on the charge $q$ due
to the magnetic moment which is proportional to the magnitude of the magnetic
moment. \ This calculation contradicts the presently accepted view that there
is no exchange of forces between a charge and a long magnet formed by stacking
magnetic moments.

\subsection{Dubious Statements in Recent Electromagnetism Textbooks}

Although there are many statements in books and articles regarding "hidden
momentum" which are of dubious validity, we wish to comment on remarks in only
two outstanding electromagnetism textbooks. \ Thus in a fine undergraduate
text on electrodynamics,\cite{G357} we find the following statement leading up
to the idea of "hidden momentum": "In fact, if the center of mass of a
localized system is at rest, its total momentum \textit{must} be zero." \ This
statement needs a qualification. \ In our system of a point charge outside a
solenoid (which can be taken of finite length) with constant currents, we
found that the center of energy\cite{Energy} was not changing and yet there
was nonzero total momentum in the system given by Eq. (1) (or by Eq. (10)).
\ This contradicts the statement of the textbook. \ If we refer back to
equation (4) above, we see that one requires a condition that no power is
introduced by external forces and transferred through space, in addition to
the condition on the center of energy [center of mass]. \ Indeed, in our
system of a point charge outside a magnetic moment where the particle in
circular motion is allowed to change speed, there is no local power delivered
by the nonelectromagnetic external (centripetal) forces, and so indeed the
total system momentum vanishes, the electromagnetic field momentum being equal
in magnitude and opposite in sign to the "hidden mechanical momentum." \ In
this case, there is work done on the charge $e$ by the electric force of the
charge $q$; however, both the charge $q$ and the charge $e$ are included
within the system and so the work done by the electric force simply transfers
energy within the system. \ 

In the same undergraduate electromagnetism textbook, there is a
calculation\cite{G521} of "hidden momentum" analogous to that given here in
Section IIIB. \ The statement appears: "Thus a magnetic dipole in an electric
field carries linear momentum, \textit{even though it is not moving!} \ This
so-called \textbf{hidden momentum} is strictly relativistic, and purely
mechanical; it precisely cancels the electromagnetic momentum stored in the
fields ..."\cite{G521s} \ The idea of the "hidden mechanical momentum"
canceling the electromagnetic field momentum is made twice in the
text.\cite{G357521} \ It is not at all clear that the momentum of the example
is actually "mechanical" if there are a large number of charges moving around
the rigid circuit, as shown in the figure of the text; closely-spaced charges
would not accelerate as isolated charges and might well carry some of the
momentum as electromagnetic momentum associated with electrostatic fields
between the charges. \ Furthermore, the "hidden mechanical momentum" does not
in any sense "cancel" the electromagnetic field angular momentum whose density
is closely related to that of the field linear momentum. \ In addition, the
example in the text seems incomplete since there is no mention of the forces
of constraint required to produce the "hidden mechanical momentum." \ Finally,
the textbook's example seems misleading since there is no mention of the
electric dipole moment (analogous to that discussed above in Section IIIE and
obviously present in the figure of the text) which must develop in the
textbook's model for a magnetic moment with "hidden mechanical momentum."
\ The electric forces associated with this electric dipole are of zero order
in $1/c^{2}$ and might dominate the order-$1/c^{2}$ considerations related to
"hidden mechanical momentum."

The leading graduate textbook of classical electromagnetism also makes remarks
regarding "hidden momentum" which must be regarded with suspicion. \ The text
states,\cite{J189} "This force [$\mathbf{F=\nabla(m\cdot B)}$] represents the
rate of change of the total mechanical momentum, including the 'hidden
momentum' associated with the presence of electromagnetic momentum." \ This
sentence might suggest that "hidden momentum" is always associated with
electromagnetic momentum. \ This is certainly not the case. \ "Hidden
mechanical momentum" requires cohesive forces within the magnetic moment which
\textit{respond} to a perturbing force in a very specific way. \ If the
cohesive forces respond so as to keep the speeds constant as in our Part II
above, then there is no "hidden mechanical momentum," nor is there any in the
more-believable case of a hydrogen-atom magnetic moment, although both of
these models do included electromagnetic field linear momentum. \ Only if the
nonelectromagnetic external forces \textit{respond} so as to keep the orbit
fixed in shape through order $1/c^{2}$ and arrange changes in particle speeds
according to one-particle energy conservation (as in our one-particle examples
of Part III) do we find "hidden mechanical momentum." \ Calculations of
"hidden mechanical momentum" of this sort appear in the problems of this
graduate text,\cite{J286618} but there is no mention of the required external
forces of constraint, or of multiparticle interactions, or of the electric
dipole moment (of the sort appearing in our Eq. (67)) which must be associated
with this kind of motion. \ Furthermore, the text continues, "The effective
force in Newton's equation of motion of mass times acceleration is
[$\mathbf{F=\nabla(m\cdot B)}$], augmented by $(1/c^{2})(d/dt)(\mathbf{E\times
m)}$, where $\mathbf{E}$ is the external electric field at the position of the
dipole."\cite{J189} \ This statement seems of dubious validity because it
assumes the existence of cohesive forces responding in such a fashion as to
give "hidden mechanical momentum" according to the currently accept view.
\ The statement takes no account of the possibility of nonmechanical momentum,
nor of the angular momentum balance, nor of the associated electric dipole
moment. \ The statement of the text arises without adequate explanation; it
merely echoes part of the presently accepted view which claims that there is
no exchange of forces between charges and long magnets.

\subsection{Understanding the Physics of the Experiments}

Understanding of the interaction of charges and magnets is fundamental to the
interpretation of the experimentally observed Aharonov-Bohm and
Aharonov-Casher phase shifts. \ Do these phase shifts represent a new
phenomenon with no classical analogue or are they the result of classical
electromagnetic interactions? \ Most of the treatments of "hidden momentum"
have been in the context of discussions claiming that the observed phase
shifts can not possibly arise based upon classical electromagnetic
interactions, and this point of view has been maintained recently with such
dogmatic certainty that any suggestion to the contrary has been rejected by
the referees and editors at the leading physics journals. \ However, now a new
set of experiments is being undertaken\cite{Bate1} which should explore new
aspects of the phase shifts, and so the theoretical classical electromagnetic
aspects are being explored anew.\cite{Bate2} \ In order to interpret the new
(and old) experimental results accurately, it is important that the errors and
uncertainties in the theoretical literature be recognized as such. \ I believe
that the idea of "hidden mechanical momentum," which now appears in the
electromagnetism textbooks, may represent a misleading distraction regarding
an interaction which is still not properly understood. \ \ The subject should
be described not as "hidden mechanical momentum" but rather as "hidden
mechanical momentum due to hidden nonelectromagnetic forces." \ Although the
idea of "hidden mechanical momentum" certainly allows curious calculations in
textbooks, it may be irrelevant to our efforts to describe nature.


\begin{thebibliography}{99}
\bibitem {AR3}This view is presented emphatically in ref. 10.
\end{thebibliography}
\end{document}